\documentclass{eas}
\usepackage{graphicx}
\begin{document}


\title{Atmospheric fluctuations below 0.1 Hz during drift-scan solar diameter measurements} 

\runningtitle{Slow seeing and drift-scanned solar diameter}

\author{C. Sigismondi}
\address{Sapienza University of Rome and ICRA, University of Nice-Sophia Antipolis, IRSOL and GPA-Observatorio Nacional Rio de Janeiro. email: sigismondi@icra.it}
\author{A. Raponi}\author{G. De Rosi}
\address{Sapienza University of Rome, Italy}
\author{M. Bianda}\author{R. Ramelli}
\address{IRSOL, Locarno, Switzerland}
\author{M. Caccia}\author{M. Maspero}\author{L. Negrini}

\address{Como University, Italy}

\author{X. Wang}

\address{Huairou Solar Observing Station, National Astronomical Observatories, Chinese Academy of Sciences, Beijing, China} 

\begin{abstract}

Measurements of the power spectrum of the seeing in the range 0.001-1 Hz have been performed in order to understand the criticity of the transits' method for solar diameter monitoring.
\end{abstract}

\maketitle


\section{Introduction}

We first measured the daytime seeing by projection of the solar image on a regular grid during a drift-scan observation.[\cite{Sigismondi}] The transit of the solar limbs above the grid are recorded, and the time intervals required to cover the evenly spaced intervals of the grid are measured by a frame by frame inspection of the video.
The standard deviation of these time intervals $\sigma$ [s] is related to the seeing $\rho$ [arcsec] by the approximate formula: $\rho = \sigma \cdot 15\cdot \cos(\delta_{\odot})$ where $\delta_{\odot}$ is the declination of the Sun at the moment of the observation.
The frequency of the seeing depends on duration of these time intervals: we used 1, 2, 4 and 8 times the single space, to sample the first 4 frequencies at the 6.3 cm Lucernaria lenses in Santa Maria degli Angeli in Rome:[\cite{Cuevas}] they are fixed in the dome and they have a focal lenght of 20 m.

These measurements have been extended to more frequencies at the 45 cm telescope of IRSOL and at the 10 cm Huariou solar station's SMAT telescope.

\begin{figure}
\centerline{\includegraphics[width=0.8\textwidth,clip=]{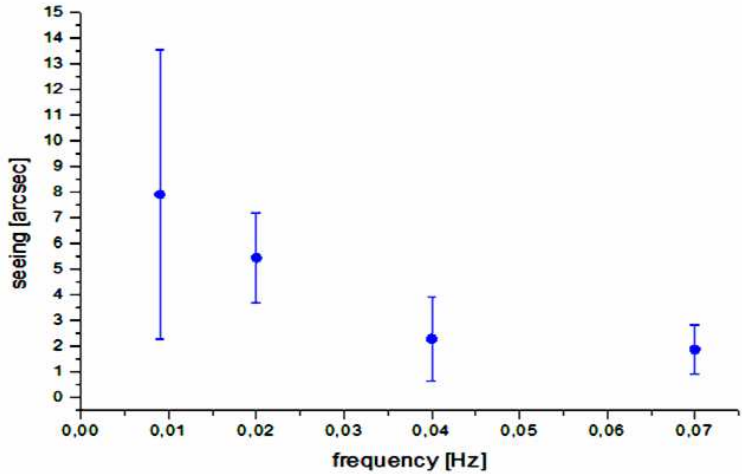}}
\caption{Seeing amplitude at different frequencies with Lucernaria lenses.}
\label{Fig. 1}
\end{figure}

\section{Image motion and drift-scan transits}

The role of seeing fluctuations between 1/10 and 1/100 of Hertz is crucial in drift-scan measurements of the solar diameter, either meridian transits or almucantarat transits. This study firstly evidenced this effect in a clear way.
 
The seeing effect results in blurring + image deformation + image motion. In our case the effect has been detected by fitting an arch of circle to a $60^o$ sector of the solar limb: the image motion concerns the whole figure of the Sun.
For Lucernaria lenses the diffraction limited the detectable amplitude of the seeing to 2.3 arcsec for $\lambda = 550$nm, while at IRSOL and at Huariou the observed fluctuations are real since this limit is respectively 0.2 and 1 arcsec.
 
It follows that during a drift-scan the timing of the transit is affected by a movement of the whole image which add itself to the drift.
 
A full Sun imager with a field of view of $4\div5$ solar radii, always in the drift-scan mode, has to monitor that slow image motion of the whole disk. The residual of this global motion with respect to the unperturbed, theoretical motion, integrated over the drift time, will give the correction to the drift time of the solar disk across a fixed hourly or altitude circle in the sky.

\begin{figure}
\centerline{\includegraphics[width=0.9\textwidth,clip=]{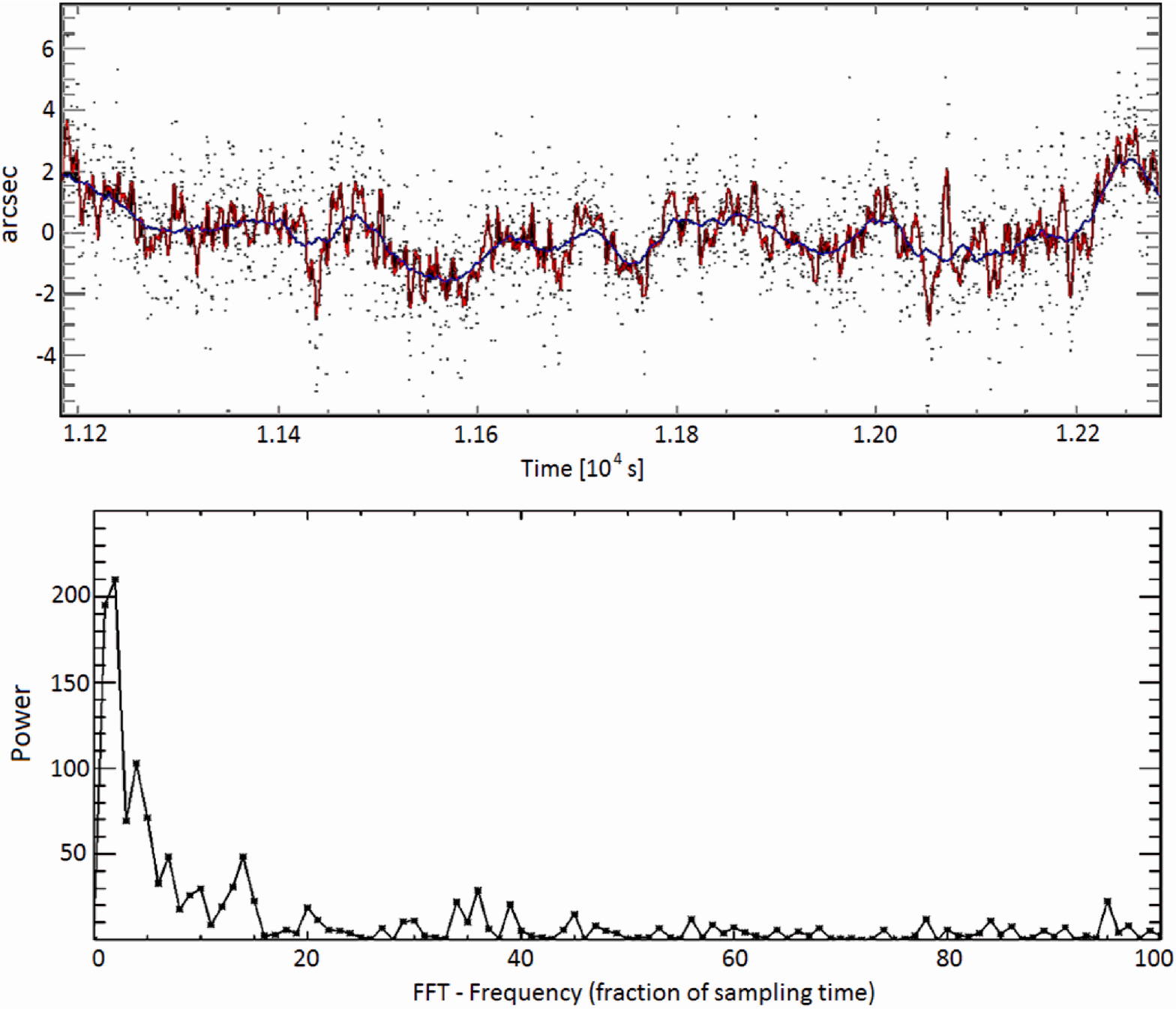}}
\caption{The Northern solar limb observed during one hour at IRSOL. The fluctuations of the center of the Sun have been recovered as ranging more than 1 arcsec.}
\label{Fig. 2}
\end{figure}

\section{Conclusions}

The difference between two following measurements of the solar diameter always experienced with drift-scan method,[\cite{Wittmann}] can be explained by the identification of this slow image motion of the whole solar disk with 1 arcsec amplitude.
This can also explain the need of averaging over several measurements to give a reference value for the solar diameter in a given day. But different meteorological conditions, even in the same day, occur during the measurements and they cannot be considered as statistically independent, as the Gaussian hypothesis requires. 
That is one of the reasons to explain the great scatter of the yearly averages published from Greenwich and Capitol Observatory.[\cite{Gething}]

\thanks{The authors C. S., M. M. and L. N. {\bf acknowledge} a grant of FRAEN Corporation to the Clavius projet for the accurate measurement of the solar diameter.}

%

\end{document}